\documentclass[onecolumn,floatfix,prd,aps,10pt]{revtex4-2}
\usepackage[applemac]{inputenc}
\usepackage[T1]{fontenc}
\pdfoutput=1
\usepackage{enumerate}
\usepackage{braket}
\usepackage{amsmath}
\usepackage{subfigure}
\usepackage{lipsum}
\usepackage{amsfonts}
\usepackage{amssymb, mathrsfs}
\usepackage{bm}
\usepackage[pdftex]{graphicx}
\usepackage{footnote}
\usepackage{hyperref}  
\hypersetup{colorlinks=true,allcolors=blue}
\usepackage{hypcap}   
\usepackage{bookmark}
\usepackage{dsfont,mathrsfs,color,url,verbatim,booktabs}
\usepackage{booktabs}
\usepackage{tabularx}
\usepackage{threeparttable}

\def\beq{\begin{equation}}
	\def\eeq{\end{equation}}
\def\bsp{\begin{split}}
	\def\esp{\end{split}}
\def\bea{\begin{eqnarray}}
	\def\eea{\end{eqnarray}}
\def\ba{\begin{array}}
	\def\ea{\end{array}}

\def\l.{\left.}
\def\r.{\right.}

\def\part{\partial}

\begin{document}
	
	\title{Static Spherically Symmetric Chaplygin and Polytropic Fluid Reconstruction in Covariant Teleparallel $F(T)$ Gravity}
	\author{A. Landry}
	\email{a.landry@dal.ca}
	\affiliation{Department of Mathematics and Statistics, Dalhousie University, Halifax, Nova Scotia, Canada, B3H 3J5}

\begin{abstract}
	We investigate static, spherically symmetric (SS) spacetimes in covariant teleparallel $F(T)$ gravity sourced by nonlinear Chaplygin and polytropic fluids. Using the covariant coframe/spin-connection (CSC) formalism, we derive the corresponding field equations and conservation laws governing admissible matter distributions and nonlinear torsion sectors. A general reconstruction procedure is developed, allowing the systematic determination of teleparallel $F(T)$ models for arbitrary coframe ansätze and fluid equations of state. Focusing on power-law configurations, we obtain several classes of reconstructed solution branches, including constant-radius, compact-object-like, and wormhole-like (WH-like) branches. The Chaplygin sector naturally leads to effective dark-energy-like and exotic-matter candidate solution branches within the reconstruction framework, which may provide admissible sectors for wormhole-like reconstructed geometries, while the polytropic sector provides reconstructed branches that may serve as physically motivated candidates for future stellar-interior and compact-object models. We discuss the associated candidate horizon and throat conditions, torsion singularities, energy conditions, and local viability properties of the reconstructed branches. The resulting geometries are organized within a teleparallel invariant classification framework, highlighting the role of nonlinear torsion corrections in shaping the solution space. Overall, this work provides a unified covariant reconstruction framework for nonlinear-fluid sectors in teleparallel $F(T)$ gravity, identifying solution branches that may serve as candidates for future compact-object, stellar-interior, and wormhole studies.
\end{abstract}

\maketitle

\section{Introduction}

The teleparallel formulation of gravity provides an alternative geometrical description of the gravitational interaction in which torsion, rather than curvature, plays the fundamental role. In this framework, gravity is described by a tetrad or coframe field together with a flat spin connection, leading to the Teleparallel Equivalent of General Relativity (TEGR), which is dynamically equivalent to Einstein's theory \cite{Aldrovandi2013,Hayashi1979,Maluf2013,Arcos2004}. The torsion scalar $T$ replaces the Ricci scalar $R$ as the gravitational Lagrangian, thereby providing a geometrically distinct but dynamically equivalent formulation of standard General Relativity in the TEGR limit.

A natural extension of TEGR is obtained by promoting the torsion scalar
to an arbitrary function $F(T)$, giving rise to modified teleparallel gravity
\cite{Ferraro2007,Linder2010,Bengochea2009}. Beyond standard $F(T)$
models, several broader classes of modified teleparallel theories have
also been proposed and investigated in recent years
\cite{Bahamonde2019,Bahamonde2023b}. These theories have attracted considerable attention as possible geometric extensions of General Relativity, particularly in cosmology, where nonlinear torsion corrections may account for the accelerated expansion of the Universe and mimic dynamical dark-energy behaviour \cite{Bahamonde2023b,Cai2016,Capozziello2011,Dent2011,Copeland2006}. Observational constraints, cosmological perturbations, and large-scale structure formation in $F(T)$ gravity have also been widely investigated \cite{Wu2010,Li2011,Izumi2013,Nesseris2013}.

The physical consistency of $F(T)$ gravity requires a covariant formulation. Early non-covariant approaches suffered from local Lorentz invariance problems, since the choice of tetrad could introduce spurious frame-dependent constraints. This issue is resolved by the covariant coframe/spin-connection (CSC) formalism, in which the tetrad $h^a{}_\mu$ and the flat spin connection $\omega^a{}_{b\mu}$ are treated together as the fundamental geometrical structure \cite{Krssak2019,Krssak2015,Golovnev2021,Bahamonde2023b}. This framework consistently separates inertial and gravitational effects and provides the natural setting for constructing physically admissible solutions in nonlinear teleparallel gravity.

Beyond cosmology, the construction of exact static and spherically symmetric (SS) solutions is an important problem in $F(T)$ gravity. Unlike General Relativity, where Birkhoff's theorem strongly restricts the SS solution space, modified teleparallel theories admit a richer set of inequivalent branches. Static SS geometries, anisotropic matter distributions, relativistic stars, and compact-object configurations have been investigated in several contexts \cite{DeBenedictis2022,Daouda2012,Wang2011,Boehmer2011}. Recent work has established covariant reconstruction procedures for
static spherically symmetric teleparallel geometries sourced by
perfect fluids and scalar fields, together with invariant
classification schemes for the resulting solution spaces
\cite{Landry2024_fluid,Landry2025_scalar,Landry2024_spherical,Landry2026_electro,roberthudsonSSpaper}.
A central objective in modified teleparallel gravity is therefore the
reconstruction of admissible functions $F(T)$ from prescribed matter
sources and geometrical ansätze. The present work extends this program to nonlinear Chaplygin and polytropic matter sectors. 

An additional tool for organizing these solutions is the invariant classification program for teleparallel geometries. Inspired by Cartan equivalence methods, this approach uses torsion invariants and Cartan scalars to distinguish inequivalent teleparallel spacetimes in a coordinate- and frame-independent manner \cite{Coley2020,McNutt2023,Coley2024,Olver1995}. Such methods are particularly relevant in nonlinear $F(T)$ gravity, where different choices of $F(T)$ may generate distinct invariant branches even when the metric ansatz has the same symmetry structure. This invariant point of view has already proved useful in the classification of static SS and generalized teleparallel de Sitter geometries \cite{Landry2024_spherical,TdSpaper}.

In the present work, we focus on nonlinear matter sources described by Chaplygin and polytropic fluids. Chaplygin-type fluids were introduced as effective dark-energy and unified dark-matter--dark-energy candidates, and they naturally generate negative pressure through equations of state of the form $p=-A/\rho^\alpha$ \cite{Kamenshchik2001,Bento2002,Bilic2002,Benaoum2002}. These properties make them especially relevant for effective dark-energy sectors and exotic-matter-supported geometries. In particular, Chaplygin fluids have been used to construct traversable wormhole configurations and generalized exotic-matter sources \cite{Lobo2006,Kuhfittig2009,MorrisThorne1988,Visser1995,Lobo2017}. Among the most intriguing strong-field applications are traversable
wormhole geometries, which provide a natural testing ground for the
interplay between exotic matter sources and effective torsion
contributions. Chaplygin fluids are particularly attractive in this context because their negative-pressure equation of state can provide NEC-violating sectors required in many wormhole constructions.

Polytropic fluids, on the other hand, provide a standard and physically motivated description of stellar interiors, relativistic fluid spheres, and compact astrophysical objects. Their equation of state, $p=K\rho^\Gamma$, has long been used in the modeling of self-gravitating fluid configurations and compact stars \cite{Tooper1964,Tooper1965,Chandrasekhar1939}. Extensions involving anisotropic polytropes and modified gravity sources further enrich the possible compact-object phenomenology \cite{Herrera2013,Harko2011}. Thus, Chaplygin and polytropic fluids offer complementary physical regimes: the former naturally probes exotic or dark-energy-like behaviour, while the latter is suited to regular matter-supported compact configurations.

While perfect-fluid, scalar-field, and electromagnetic matter sectors have already been investigated within static spherically symmetric teleparallel gravity, a systematic covariant treatment of nonlinear Chaplygin and polytropic equations of state remains incomplete. The interplay between nonlinear matter sectors, torsion reconstruction, and invariant classification therefore remains largely unexplored.

Motivated by these considerations, the goal of this work is to develop a covariant analysis of static SS solutions in teleparallel $F(T)$ gravity sourced by Chaplygin and polytropic fluids. We adopt the CSC formalism to ensure local Lorentz covariance, derive the corresponding field equations and conservation laws, and construct a reconstruction procedure for admissible nonlinear $F(T)$ models. Particular attention is devoted to constant-radius and areal-radius reconstruction branches, including compact-object-like and wormhole-like sectors, as well as to their torsion invariants, energy conditions, and local viability conditions. From a broader perspective, the reconstructed solution space may also be interpreted within the framework of dynamical systems and invariant classification techniques, which provide complementary tools for understanding the global structure of nonlinear teleparallel models \cite{Coley2009,Coley2020,McNutt2023,Coley2024}.

The main contributions of the present work are:
\begin{itemize}
	\item We derive the covariant static SS field equations for teleparallel $F(T)$ gravity coupled to nonlinear Chaplygin and polytropic fluids.
	\item We obtain the corresponding conservation laws and show how they replace the matter-sector constraints used in other matter models.
	\item We construct reconstruction equations for admissible nonlinear $F(T)$ functions associated with power-law coframe ans\"atze.
	\item We analyze the conditions under which reconstructed branches may exhibit wormhole-like behaviour.
	\item We investigate the radial null energy condition and discuss how effective NEC violation may be transferred from the physical fluid sector to the effective torsion sector.
	\item We organize the resulting branches within the teleparallel invariant classification framework.
	\item We provide a unified reconstruction framework encompassing representative ordinary, dark-energy-like, and exotic matter sectors within a common covariant teleparallel setting.
\end{itemize}

From a physical perspective, these reconstructed branches provide a unified covariant framework for studying nonlinear-fluid compact objects, effective dark-energy sectors, regular-core strong-field geometries, and traversable wormholes in teleparallel $F(T)$ gravity. The reconstructed models may also provide useful phenomenological extensions of previous perfect-fluid, scalar-field, and wormhole solutions in modified teleparallel gravity \cite{Landry2024_fluid,Landry2025_scalar,Landry2026_electro,LandryWHmass2026}.

More broadly, the present study contributes to the growing program of
understanding how nonlinear matter sectors interact with nonlinear
torsion corrections in modified teleparallel gravity. Such analyses
may help clarify the extent to which effective gravitational degrees
of freedom can mimic exotic matter sources while preserving the
covariant structure of the theory.

The paper is organized as follows. In Section~\ref{sec:field_equations}, we introduce the covariant teleparallel $F(T)$ field equations with nonlinear fluid sources. In Section~\ref{sec:csc_pair}, we specialize to the static SS CSC pair and derive the corresponding torsion scalar. In Sections~\ref{sec:chaplygin} and~\ref{sec:polytropic}, we analyze the Chaplygin and polytropic equations of state and their conservation laws. In the following sections, we study the constant-radius and areal-radius reconstruction sectors, including compact-object-like and wormhole-like branches. Finally, we discuss local viability conditions, invariant classification, and physical implications, before summarizing our results.


\section{Teleparallel Field Equations and Geometry}\label{sec:field_equations}

\subsection{Covariant Field Equations with Nonlinear Fluids}

We consider the covariant formulation of teleparallel $F(T)$ gravity
coupled to a nonlinear fluid source. The matter sector is described by
a fluid Lagrangian $\mathcal{L}_m$:
\begin{equation}
	S=\int d^4x\,h\left[\frac{1}{2\kappa}F(T)+\mathcal{L}_m\right],
\end{equation}
where $h=\det(h^a{}_\mu)$ and $\kappa=8\pi$. For perfect and nonlinear fluid sources, different effective fluid representations may be adopted, such as $L_m=-\rho$ or $L_m=p$. Since the present analysis is formulated at the effective energy-momentum tensor level, the subsequent reconstruction procedure remains independent of this representation choice within the adopted phenomenological description. The use of different fluid Lagrangians in modified gravity theories has been discussed extensively in the literature, particularly in the context of non-minimal matter couplings and effective fluid descriptions \cite{Harko2011}.

The symmetric and antisymmetric field equations are
\begin{align}
	\kappa \Theta_{(ab)}
	=&
	F_T\mathring{G}_{ab}
	+
	F_{TT}S^\mu{}_{(ab)}\partial_\mu T
	+
	\frac{g_{ab}}{2}\left[F-TF_T\right],
\\
		0=&F_{TT}S^\mu{}_{[ab]}\partial_\mu T .
\end{align}
The antisymmetric equations vanish identically in the TEGR limit, providing an immediate consistency check of the covariant formulation. In the TEGR limit $F(T)=T+\beta$, one has $F_T=1$ and $F_{TT}=0$, and the field equations reduce to the Einstein equations with an effective cosmological constant $\Lambda_{\rm eff}=-\beta/2$. For the CSC pair used below, the antisymmetric equations act as consistency constraints on admissible branches rather than as independent dynamical equations for the matter sector.

The nonlinear fluid source is allowed to be either isotropic or
anisotropic. Its energy-momentum tensor is written as
\begin{equation}
	\Theta^\mu{}_\nu=
	\mathrm{diag}\left[-\rho(r),p_r(r),p_t(r),p_t(r)\right].
\end{equation}
The isotropic limit 
$p_r=p_t=p$ is recovered as a particular case and
will be adopted throughout most of the present work. Here \(\rho\) denotes the energy density, while \(p_r\) and \(p_t\) denote the radial and tangential pressures.

Compared with the perfect-fluid sector studied in Ref.~\cite{Landry2024_fluid},
the nonlinear equations of state considered here introduce additional
density-dependent couplings. In contrast with field-theoretic matter models
\cite{Landry2025_scalar,Landry2026_electro}, the matter dynamics is governed entirely by
the conservation law and the chosen equation of state.

\subsection{Static Spherically Symmetric CSC Pair}\label{sec:csc_pair}

We adopt the same static spherically symmetric coframe/spin-connection pair as refs. \cite{Krssak2019,Krssak2015,Landry2024_spherical,Landry2024_fluid,Landry2025_scalar,Landry2026_electro,roberthudsonSSpaper},
\begin{equation}
	h^a{}_\mu=
	\mathrm{diag}\left(A_1(r),A_2(r),A_3(r),A_3(r)\sin\theta\right),
\end{equation}
with
\begin{equation}
	\omega^{2}{}_{33}=\omega^{2}{}_{44}=\frac{\delta}{A_3(r)},
	\qquad
	\omega^{3}{}_{44}=-\frac{\cot\theta}{A_3(r)} .
\end{equation}
The metric is
\begin{equation}
	ds^2=
	-A_1^2(r)dt^2
	+
	A_2^2(r)dr^2
	+
	A_3^2(r)d\Omega^2 .
\end{equation}

The torsion scalar is written as
\begin{equation}
	T=T\left(A_1,A_2,A_3,A_1',A_3'\right),
\end{equation}
and becomes explicit once a radial gauge, such as $A_3=c_0$ or $A_3=r$, is
chosen. The torsion scalar constitutes the lowest-order scalar invariant of
the teleparallel geometry and plays a central role in the invariant
classification of static spherically symmetric CSC pairs.

At the lowest invariant orders, examples of scalar quantities useful for invariant classification include
\begin{equation}
	I_1=T,\qquad
	I_2=T_{\alpha\mu\nu}T^{\alpha\mu\nu},\qquad
	I_3=\nabla_\mu T\nabla^\mu T .
\end{equation}
These invariants do not exhaust the full Cartan invariant structure and should therefore be interpreted as scalar torsion diagnostics rather than a complete Cartan invariant basis. Their role is qualitative in the present reconstruction analysis, while a complete equivalence classification requires the full Cartan algorithm \cite{Coley2020,McNutt2023,Coley2024,Olver1995}.

\section{Conservation Law Solutions}

\subsection{Nonlinear Fluid Conservation Law}

The matter conservation equation is
\begin{equation}
	\mathring{\nabla}_\mu \Theta^{\mu}{}_\nu=0 .
\end{equation}
For a static spherically symmetric anisotropic fluid, the radial component gives
\begin{equation}
	p_r'
	+
	\frac{A_1'}{A_1}\left(\rho+p_r\right)
	+
	2\frac{A_3'}{A_3}\left(p_r-p_t\right)
	=0 .
\end{equation}

For an isotropic nonlinear fluid, $p_r=p_t=p$, this reduces to
\begin{equation}\label{eqn11}
	p'
	+
	\frac{A_1'}{A_1}\left(\rho+p\right)
	=0 .
\end{equation}
Whenever the equation of state can be inverted locally, $\rho=\rho(p)$, Eq.~\eqref{eqn11}
may be formally integrated as
\begin{equation}
	A_1(r)=A_{10}\exp\left[-\int^p
	\frac{d\bar p}{\rho(\bar p)+\bar p}\right].
\end{equation}
Equation \eqref{eqn11} provides the master conservation law governing all
nonlinear-fluid sectors considered below. Once an equation of state
$p=p(\rho)$ is specified, it yields a first-order differential equation
for the density profile. This equation provides the matter-sector constraint replacing the source-specific conservation laws used in other teleparallel models \cite{Landry2024_fluid}.

\subsection{Chaplygin Fluid Sector}\label{sec:chaplygin}

The generalized Chaplygin equation of state is
\begin{equation}
	p=-\frac{A}{\rho^\alpha},
	\qquad
	A>0,
	\qquad
	0\leq \alpha\leq 1 .
\end{equation}
A related extension is the modified Chaplygin gas,
$p=B\rho-A/\rho^\alpha$, which interpolates between a barotropic
fluid contribution and a Chaplygin-type negative-pressure sector \cite{Benaoum2002}. The standard Chaplygin gas corresponds to $\alpha=1$,
while $\alpha=0$ yields a constant negative-pressure sector.
In the high-density regime $\rho^{\alpha+1}\gg A$, the pressure
contribution becomes negligible compared with the density,
$|p|/\rho\ll 1$. Near $\rho^{\alpha+1}\simeq A$ it behaves approximately as an effective cosmological-constant sector. This behaviour explains why Chaplygin-type fluids have often been considered as unified dark-matter/dark-energy candidates in cosmology \cite{Kamenshchik2001,Bento2002,Bilic2002,Benaoum2002}.

For an isotropic Chaplygin fluid, the conservation law becomes
\begin{equation}
	-\alpha A\rho^{-\alpha-1}\rho'
	+
	\frac{A_1'}{A_1}
	\left(
	\rho-\frac{A}{\rho^\alpha}
	\right)
	=0 .
\end{equation}
Equivalently,
\begin{equation}
	\rho'
	=
	\frac{A_1'}{A_1}
	\frac{\rho\left(\rho^{\alpha+1}-A\right)}{\alpha A}.
\end{equation}
The adiabatic sound speed is
\begin{equation}
c_s^2=\frac{dp}{d\rho}=\frac{\alpha A}{\rho^{\alpha+1}}.
\end{equation}
For $0\le \alpha \le 1$ and \(\rho>0\), positivity of the adiabatic sound speed squared follows formally for the chosen parameter range. The causality condition $c_s^2\le1$ imposes additional restrictions
on the admissible density range. Therefore, viable Chaplygin branches are restricted to density ranges satisfying both positivity and subluminal propagation.

The radial null energy condition is
\begin{equation}
	\rho+p_r=
	\rho-\frac{A}{\rho^\alpha}.
\end{equation}
Thus the Chaplygin fluid violates or saturates the radial NEC whenever
\begin{equation}
	\rho^{\alpha+1}\leq A .
\end{equation}
This condition identifies admissible NEC-violating regimes for the physical fluid sector, but does not by itself guarantee the existence of a complete traversable wormhole solution. Consequently, low-density regimes may naturally favour NEC violation,
whereas sufficiently dense configurations behave more like ordinary
matter sources. This behaviour makes Chaplygin fluids particularly
attractive for connecting dark-energy-like sectors with candidate wormhole-like geometries.

\subsection{Polytropic Fluid Sector}\label{sec:polytropic}

The polytropic equation of state is
\begin{equation}
	p=K\rho^\Gamma,
	\qquad
	\Gamma=1+\frac{1}{n_p},
\end{equation}
where $K$ is the polytropic constant and $n_p$ is the polytropic index. Typical astrophysical applications involve values of \(n_p\) of order unity, with commonly used models spanning approximately \(n_p \sim 1-5\). Such polytropic models provide the basis of many classical and relativistic
stellar-structure calculations ranging from white dwarfs to neutron-star
configurations \cite{Tooper1964,Tooper1965,Chandrasekhar1939,Herrera2013}. Common astrophysical values include $n_p=1.5$ for non-relativistic
degenerate matter, $n_p=3$ for relativistic degenerate matter, and
$n_p\simeq 5$ for diffuse stellar envelopes.

For an isotropic polytropic fluid, the conservation law gives
\begin{equation}
	K\Gamma \rho^{\Gamma-1}\rho'
	+
	\frac{A_1'}{A_1}
	\left(
	\rho+K\rho^\Gamma
	\right)
	=0 .
\end{equation}
Therefore,
\begin{equation}
	\rho'
	=
	-\frac{A_1'}{A_1}
	\frac{\rho+K\rho^\Gamma}{K\Gamma\rho^{\Gamma-1}} .
\end{equation}
For a barotropic polytropic fluid, the specific enthalpy contribution is
\begin{equation}
	h(\rho)=\int \frac{dp}{\rho}
	=\frac{K\Gamma}{\Gamma-1}\rho^{\Gamma-1}.
\end{equation}
This expression corresponds to the barotropic enthalpy contribution
commonly used in stellar-structure models; a fully relativistic
definition may instead involve $\int dp/(\rho+p)$ \cite{Tooper1964,Tooper1965,Chandrasekhar1939}. It is quoted here only to connect the polytropic sector with standard stellar-structure notation.

For $K>0$ and $\rho>0$, one has
\begin{equation}
	\rho+p=\rho+K\rho^\Gamma>0 ,
\end{equation}
so the polytropic branch is naturally associated with compact-object and
stellar-interior configurations rather than exotic matter within suitable parameter ranges.

The adiabatic sound speed is
\begin{equation}
	c_s^2=\frac{dp}{d\rho}=K\Gamma\rho^{\Gamma-1}.
\end{equation}

Physical configurations generally require $c_s^2>0$ and, in relativistic
units, $c_s^2\le1$.

The Chaplygin and polytropic equations of state may be viewed as
representative examples of exotic and ordinary matter sectors,
respectively. Together they provide a convenient testing ground
for investigating how nonlinear matter sources affect torsion
reconstruction, invariant classification, and the existence of
compact-object and wormhole-like solutions in teleparallel
$F(T)$ gravity.

\begin{table}[h]
	\centering
	\caption{Comparison of the nonlinear fluid sectors considered in this work.}
	\label{tab:fluidcomparison}
	\begin{tabularx}{\textwidth}{l l l l l}
		\toprule
		Fluid sector &
		Equation of state &
		NEC behaviour &
		Adiabatic sound speed \(c_s^2=dp/d\rho\) &
		Physical interpretation \\
		\midrule
		
		Chaplygin &
		$p=-A/\rho^{\alpha}$ &
		May be violated &
		$c_s^2=\frac{\alpha A}{\rho^{\alpha+1}}$ &
		Dark-energy-like/wormhole-like sectors \\
		
		Modified Chaplygin &
		$p=B\rho-A/\rho^{\alpha}$ &
		Model dependent &
		$c_s^2=B+\frac{\alpha A}{\rho^{\alpha+1}}$ &
		Dark-fluid interpolation and regular-core models \\
		
		Polytropic &
		$p=K\rho^{\Gamma}$ &
		Usually satisfied &
		$c_s^2=K\Gamma\rho^{\Gamma-1}$ &
		Candidate compact-star and stellar-interior sectors \\
		
		TEGR fluid limit &
		$F(T)=T+\beta$ &
		Matter dependent &
		Matter dependent &
		General-relativistic fluid configurations \\
		
		\bottomrule
	\end{tabularx}
\end{table}

\begin{figure}
	\includegraphics[width=0.70\textwidth]{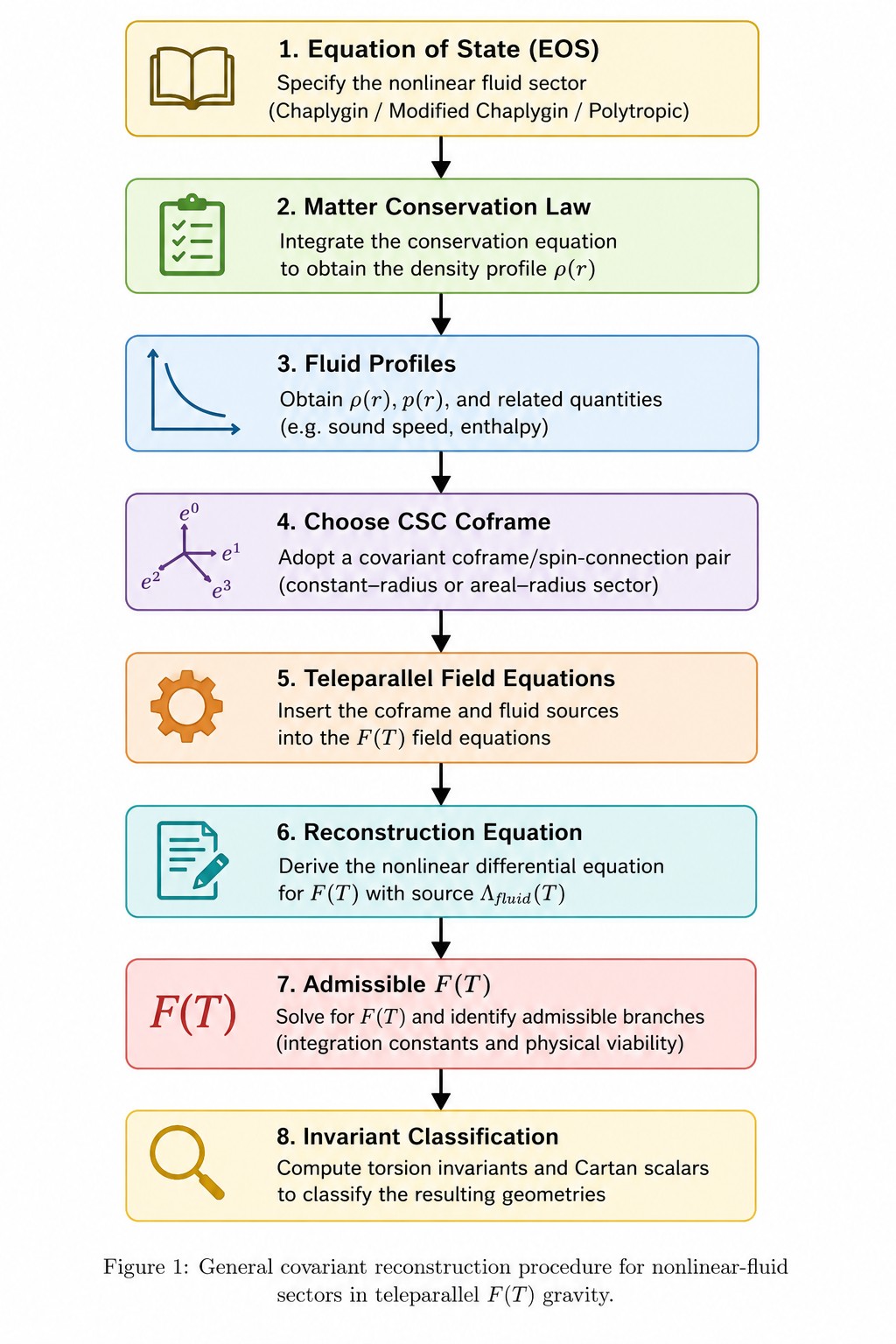}
	\label{fig1}
\end{figure}

These nonlinear equations of state therefore span two complementary regions of the teleparallel matter sector, providing representative examples of exotic and ordinary self-gravitating fluids. The main physical characteristics of the Chaplygin and polytropic branches are summarized in Table~\ref{tab:fluidcomparison}. These nonlinear fluid sectors provide complementary matter sources for the reconstruction of admissible covariant teleparallel $F(T)$ models, where the process is shown in the figure \ref{fig1}. In the following sections, they are incorporated into both constant-radius and areal-radius reconstruction schemes.

\section{General Teleparallel $F(T)$ Solutions for Non-Linear Fluids}

\subsection{Field Equations for the Constant-Radius Sector $A_3=c_0$}\label{sect7}

\begin{figure}
	\includegraphics[width=0.75\textwidth]{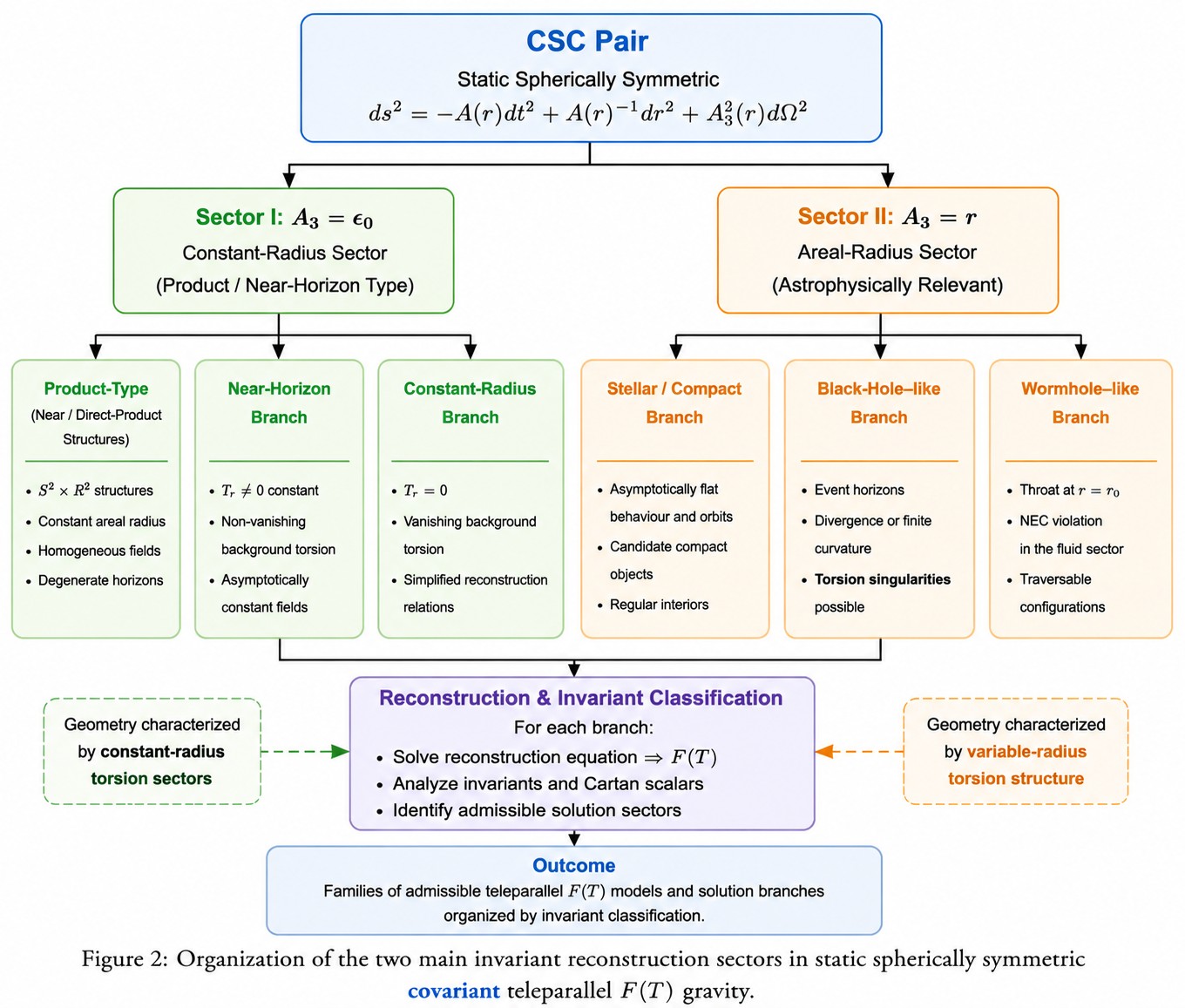}
	\label{fig2}
\end{figure}

The constant-radius sector $A_3=c_0$ corresponds to a
teleparallel analogue of Nariai- and Bertotti–Robinson-type
product geometries, where the areal radius remains fixed. Such product-type sectors are closely related to the invariant
classification of teleparallel de Sitter-like geometries
\cite{TdSpaper}. These configurations provide useful laboratories for studying torsion-dominated branches, near-horizon-like limits, and
effective compactification sectors in nonlinear $F(T)$ gravity. In contrast with the areal-radius sector, this branch does not describe
ordinary stellar interiors, but rather product-type or near-horizon
geometries \cite{Landry2024_spherical,TdSpaper}. The organization of the two reconstruction sectors is summarized in Fig.~\ref{fig2}.

For $A_3=c_0$, the reduced field equations become
\begin{align}
	\kappa\rho
	&=
	-\frac{1}{2}\left[F-TF_T\right]
	-\frac{2\delta}{A_2c_0}\partial_r(F_T)
	+\frac{F_T}{c_0^2},
	\\
	\kappa p_r
	&=
	\frac{1}{2}\left[F-TF_T\right]
	-\frac{F_T}{c_0^2},
	\\
	\kappa p_t
	&=
	-\frac{1}{2}\left[F-TF_T\right]
	-\frac{\partial_r(F_T)}{A_2}
	\left[
	\frac{\delta}{c_0}
	+
	\frac{A_1'}{A_1A_2}
	\right]
	-F_T
	\left[
	\frac{A_1''}{A_2^2A_1}
	-
	\frac{A_1'}{A_1A_2}
	\frac{A_2'}{A_2^2}
	\right].
\end{align}

The torsion scalar is
\begin{equation}
	T(r)=
	-2
	\left(
	\frac{\delta}{c_0}
	\right)
	\left(
	\frac{\delta}{c_0}
	+
	\frac{2A_1'}{A_1A_2}
	\right).
\end{equation}

For the power-law coframe
\begin{equation}
	A_1(r)=a_0r^a,
	\qquad
	A_2(r)=b_0r^b,
\end{equation}
one obtains
\begin{equation}
	T(r)=T_0+T_1r^{-(b+1)},
\end{equation}
where
\begin{equation}
	T_0=-\frac{2}{c_0^2},
	\qquad
	T_1=-\frac{4a\delta}{c_0b_0}.
\end{equation}
Since
\[
T_0=-\frac{2}{c_0^2}<0,
\]
the constant-radius branch possesses a non-vanishing
background torsion even when the power-law contribution
vanishes asymptotically.

For $b\neq -1$,
\begin{equation}
	r^{-(b+1)}=\frac{T-T_0}{T_1}.
\end{equation}

\subsection{Reconstruction in the Constant-Radius Sector}\label{sect8}

Using the inversion relation
\begin{equation}
	r^{-(b+1)}=\frac{T-T_0}{T_1},
\end{equation}
all radial functions can be rewritten as functions of $T$. The reduced field
equations then lead to the reconstruction equation
\begin{equation}\label{eqn35}
	(T-T_0)F_{TT}
	+
	\gamma(a,b)F_T
	=
	\Lambda_{\rm fluid}(T),
\end{equation}
where
\begin{equation}
	\gamma(a,b)=\frac{b+1-2a}{b+1}.
\end{equation}
A necessary consistency requirement is that the fluid source
$\Lambda_{\rm fluid}(T)$ be expressible solely as a function
of $T$, which imposes compatibility conditions between the
density profile, the conservation law, and the chosen coframe
ansatz.

The source function $\Lambda_{\rm fluid}(T)$ is determined by combining
the selected equation of state with the conservation law and the
chosen radial ansatz. For the Chaplygin case,
\begin{equation}
	p=-\frac{A}{\rho^\alpha},
\end{equation}
whereas for the polytropic case,
\begin{equation}
	p=K\rho^\Gamma .
\end{equation}

\begin{figure}
	\includegraphics[width=0.70\textwidth]{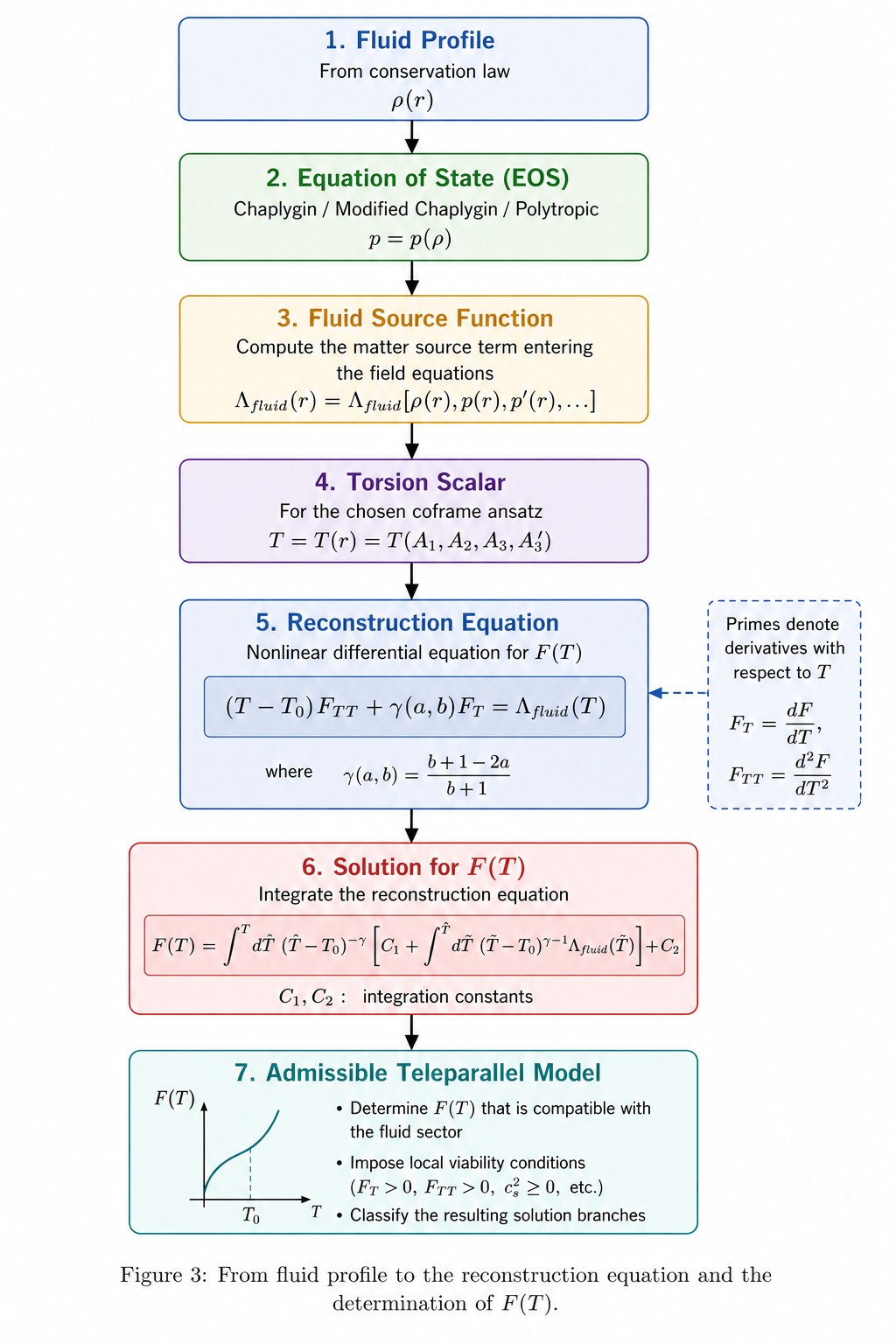}
	\label{fig3}
\end{figure}

The formal reconstruction solution is
\begin{equation}\label{eqn39}
	F(T)=
	\int^T d\bar{T}\,
	(\bar{T}-T_0)^{-\gamma}
	\left[
	C_1+
	\int^{\bar{T}}d\tilde{T}\,
	(\tilde{T}-T_0)^{\gamma-1}
	\Lambda_{\rm fluid}(\tilde{T})
	\right]
	+C_2 .
\end{equation}
Equation \eqref{eqn39} shows that the reconstruction equation remains linear in the unknown function $F(T)$ despite the nonlinear character of the matter sector. The nonlinearity of the matter sector is transferred into the effective reconstruction source$\Lambda_{fluid}(T)$, while the differential operator acting on F(T) remains linear. The integration constants $C_1$ and $C_2$ parametrize the homogeneous reconstruction sector and possible constant shifts of the reconstructed gravitational sector. Different choices of the coframe ansatz or fluid profile may lead to different reconstructed functions $F(T)$, illustrating the non-uniqueness of the inverse reconstruction problem. Reconstruction procedures of this type are common in modified teleparallel gravity and provide a useful bridge between prescribed matter sectors and admissible torsion functions \cite{Bahamonde2019,Bahamonde2023b,Cai2016,Landry2024_fluid,Landry2025_scalar,Landry2026_electro}. The TEGR limit is discussed separately below as a consistency check. The reconstruction workflow is summarized in Fig.~\ref{fig3}.

\subsection{TEGR Consistency Checks}

An important consistency requirement for any reconstructed teleparallel model is the recovery of the Teleparallel Equivalent of General Relativity (TEGR). In this limit,

\begin{equation}
	F(T)=T+\beta,
\end{equation}

with

\begin{equation}
	F_T=1,
	\qquad
	F_{TT}=0.
\end{equation}

The nonlinear torsion contributions therefore vanish and the field equations reduce to those of General Relativity with an effective cosmological constant

\begin{equation}
	\Lambda_{\rm eff}=-\frac{\beta}{2},
\end{equation}

written in the covariant coframe/spin-connection formalism
\cite{Aldrovandi2013,Maluf2013,Krssak2015}.

Substituting \(F_{TT}=0\) into Eq. ~\eqref{eqn35} yields, within this reduced reconstruction sector, the TEGR compatibility constraint within the reduced reconstruction equation
\begin{equation}
	\Lambda_{\rm fluid}(T)=0,
\end{equation}
which becomes an algebraic compatibility condition relating the fluid sector and the effective cosmological constant contribution. This condition applies only to the homogeneous reconstruction equation considered in the constant-radius power-law sector and should not be interpreted as a universal requirement for arbitrary TEGR fluid solutions.

Consequently, both the constant-radius and areal-radius branches continuously recover the corresponding Einstein-sector solutions. The TEGR limit therefore provides a useful consistency check on the reconstruction procedure, with all deviations from General Relativity encoded in the nonlinear torsion corrections generated by $F_{TT}\neq 0$ \cite{Aldrovandi2013,Maluf2013,Krssak2015}. This demonstrates that the reconstruction procedure admits an appropriate Einsteinian limit whenever the nonlinear equation of state is compatible with the TEGR branch.

\subsection{Field Equations for the Areal-Radius Sector $A_3=r$}

The gauge choice $A_3=r$ corresponds to the standard
areal-radius coordinate and provides the natural framework
for describing stellar, black-hole-like, and wormhole-like
configurations \cite{DeBenedictis2022,Daouda2012,Wang2011,Boehmer2011}. This sector will therefore be the most relevant one for compact-object and wormhole applications. For $A_3=r$, the reduced field equations are
\begin{align}
	\kappa\rho
	&=
	-\frac{1}{2}\left[F-TF_T\right]
	-\frac{2}{A_2}\partial_r(F_T)
	\left(
	\frac{\delta}{r}
	+
	\frac{1}{A_2r}
	\right)
	+
	F_T
	\left[
	2\frac{A_1'}{A_1A_2}\frac{1}{A_2r}
	-\frac{1}{A_2^2r^2}
	+\frac{1}{r^2}
	\right],
	\\
	\kappa p_r
	&=
	\frac{1}{2}\left[F-TF_T\right]
	+
	F_T
	\left[
	2\frac{A_1'}{A_1A_2}\frac{1}{A_2r}
	+
	\frac{1}{A_2^2r^2}
	-\frac{1}{r^2}
	\right],
	\\
	\kappa p_t
	&=
	-\frac{1}{2}\left[F-TF_T\right]
	-\frac{\partial_r(F_T)}{A_2}
	\left[
	\frac{\delta}{r}
	+
	\frac{A_1'}{A_1A_2}
	+
	\frac{1}{A_2r}
	\right]
	-
	F_T
	\left[
	\frac{A_1''}{A_2^2A_1}
	-
	\frac{A_1'}{A_1A_2}
	\frac{A_2'}{A_2^2}
	+
	\frac{A_1'}{A_1A_2}
	\frac{1}{A_2r}
	-
	\frac{1}{A_2r}
	\frac{A_2'}{A_2^2}
	\right].
\end{align}

The torsion scalar is
\begin{equation}
	T(r)=
	-2
	\left(
	\frac{\delta}{r}
	+
	\frac{1}{A_2r}
	\right)
	\left(
	\frac{\delta}{r}
	+
	\frac{1}{A_2r}
	+
	\frac{2A_1'}{A_1A_2}
	\right).
\end{equation}
The resulting torsion scalar and its derivatives generate the
lowest-order invariant hierarchy used in the teleparallel invariant-based
classification program, allowing different reconstructed
branches to be distinguished independently of coordinate
choices \cite{Coley2020,McNutt2023,Coley2024,Landry2024_spherical}.

\subsection{Power-Law Reconstruction for $A_3=r$}\label{sect10}

For
\begin{equation}
	A_1(r)=a_0r^a,
	\qquad
	A_2(r)=b_0r^b,
\end{equation}
the torsion scalar takes the multi-scale form
\begin{equation}\label{eqn45}
	T(r)=T_0r^{-2}+T_1r^{-(b+2)}+T_2r^{-2(b+1)},
\end{equation}
where
\begin{equation}
	T_0=-2,
	\qquad
	T_1=\frac{4(1+a)}{b_0},
	\qquad
	T_2=-\frac{2(1+a)^2}{b_0^2}.
\end{equation}
Depending on the value of $b$, the three contributions in Eq.~\eqref{eqn45}
may dominate in different radial regimes. The $T_0r^{-2}$ term
represents the universal areal-radius contribution, whereas the
$T_1r^{-(b+2)}$ and $T_2r^{-2(b+1)}$ terms encode coframe-dependent
torsion corrections.

This multi-scale structure generates composite reconstructed models such as
\begin{equation}\label{eqn47}
	F(T)=
	\alpha(T-T_\star)^{n_1}
	+
	\beta(T-T_\star)^{n_2}
	+
	\gamma,
\end{equation}
with
\begin{equation}\label{eqn48}
	n_1=\frac{2a}{b+2},
	\qquad
	n_2=\frac{a}{b+1}.
\end{equation}

Log-corrected branches may be written as
\begin{equation}\label{eqn49}
	F(T)=
	\alpha(T-T_\star)^n
	\left[
	1+\eta\ln(T-T_\star)
	\right]
	+\beta T,
\end{equation}
while exponential-power hybrid models take the form
\begin{equation}\label{eqn50}
	F(T)=
	\alpha(T-T_\star)^n+\beta e^{\lambda T}.
\end{equation}
These classes include several functional forms that have been explored in cosmological and strong-field applications of teleparallel gravity \cite{Ferraro2007,Linder2010,Bengochea2009,Bahamonde2019,Bahamonde2023b,Cai2016}. Consequently, the present reconstruction scheme may generate a broad class of phenomenologically relevant teleparallel theories.

Singularities may occur not only through metric functions but also through
divergences of torsion invariants such as $T$, $T_{\alpha\mu\nu}T^{\alpha\mu\nu}$,
or higher-order invariant combinations \cite{Coley2020,McNutt2023,Coley2024,Landry2024_spherical}. Consequently, a metric-regular configuration need not be torsion-regular, and vice versa.

\begin{table}[ht]
	\centering
	\caption{Examples of reconstructed teleparallel models.}
	\label{tab:reconstructedmodels}
	\begin{tabularx}{\textwidth}{ccc}
		\toprule
		Model & Form & Role in reconstruction \\
		\midrule
		TEGR & $T+\beta$ & GR limit \\
	Power law & $T+\alpha T^n$ or $\alpha(T-T_\star)^n+\beta$ & torsion correction \\
		Shifted power law & $\alpha(T-T_\star)^n+\beta$ & reconstructed branch \\
		Log corrected & $\alpha(T-T_\star)^n[1+\eta\ln(T-T_\star)]+\beta T$ & logarithmic correction \\
		Exponential-power & $\alpha(T-T_\star)^n+\beta e^{\lambda T}$ & candidate strong-field modification \\
		\bottomrule
	\end{tabularx}
\end{table}

These reconstructed model classes will now be combined with the
Chaplygin and polytropic equations of state in order to obtain reconstructed nonlinear-fluid source branches.

\section{Chaplygin and Polytropic Solutions, Stability and Classification}

\subsection{Chaplygin and Polytropic Reconstructed Nonlinear-Fluid Source Branches}

The methodology adopted throughout the paper can be summarized as follows: (i) specify the nonlinear equation of state; (ii) integrate the conservation law; (iii) choose a covariant CSC coframe; (iv) derive the reconstruction equation; (v) determine the admissible $F(T)$ solution; and (vi) classify the resulting branches according to their invariant torsion structure.

We now combine the reconstructed teleparallel models of Sections \ref{sect7}--\ref{sect10}
with the nonlinear equations of state introduced in Sections \ref{sec:chaplygin}--\ref{sec:polytropic}.
This yields reconstructed nonlinear-fluid source branches in which the density profile,
radial pressure, and admissible function $F(T)$ are constrained
simultaneously. 

For the Chaplygin branch,
\begin{equation}
	p_r=-\frac{A}{\rho^\alpha},
\end{equation}
the radial field equation becomes
\begin{equation}\label{eqn52}
	-\kappa\frac{A}{\rho^\alpha}
	=
	\frac{1}{2}\left[F-TF_T\right]
	+
	F_T\mathcal{P}_r(r),
\end{equation}
where
\begin{equation}
	\mathcal{P}_r(r)
	=
	2\frac{A_1'}{A_1A_2}\frac{1}{A_2r}
	+
	\frac{1}{A_2^2r^2}
	-
	\frac{1}{r^2}.
\end{equation}

For the modified Chaplygin branch,
\begin{equation}
	p_r=B\rho-\frac{A}{\rho^\alpha},
\end{equation}
the radial field equation becomes
\begin{equation}
	\kappa\left(B\rho-\frac{A}{\rho^\alpha}\right)
	=
	\frac{1}{2}\left[F-TF_T\right]+F_T\mathcal{P}_r(r).
\end{equation}

For the polytropic branch,
\begin{equation}
	p_r=K\rho^\Gamma,
\end{equation}
one obtains
\begin{equation}
	\kappa K\rho^\Gamma
	=
	\frac{1}{2}\left[F-TF_T\right]
	+
	F_T\mathcal{P}_r(r).
\end{equation}

The density equation is obtained from the temporal field equation,
\begin{equation}
	\kappa\rho
	=
	-\frac{1}{2}\left[F-TF_T\right]
	+
	\mathcal{D}_{\rho}[F_T]
	+
	F_T\mathcal{P}_{\rho}(r),
\end{equation}
where
\begin{equation}
	\mathcal{D}_{\rho}[F_T]
	=
	-\frac{2}{A_2}\partial_r(F_T)
	\left(
	\frac{\delta}{r}
	+
	\frac{1}{A_2r}
	\right),
\end{equation}
and
\begin{equation}\label{eqn58}
	\mathcal{P}_{\rho}(r)
	=
	2\frac{A_1'}{A_1A_2}\frac{1}{A_2r}
	-
	\frac{1}{A_2^2r^2}
	+
	\frac{1}{r^2}.
\end{equation}
Equations \eqref{eqn52}--\eqref{eqn58}, together with the conservation law, form a closed reconstruction system for $\{F(T),\rho(r)\}$ once the coframe functions $A_1(r)$ and $A_2(r)$ are specified \cite{Landry2024_fluid,Landry2025_scalar,Landry2026_electro}. The difference between the branches is therefore encoded in the nonlinear relation between \(p_r\) and \(\rho\), while the geometric operators \(\mathcal{P}_r\), \(\mathcal{P}_\rho\), and \(\mathcal{D}_\rho\) are fixed by the chosen CSC sector.

Certain reconstructed Chaplygin branches may admit regular-core behaviour
in which the density remains finite while the sector
approaches a de Sitter-like effective geometric regime \cite{TdSpaper}. In the TEGR limit, the reconstructed branches reduce to the corresponding General Relativity fluid configurations whenever compatible solutions exist. The nonlinear torsion sector therefore acts as a deformation of the underlying Einstein-fluid branch rather than an entirely disconnected solution family. Consequently, both Chaplygin and polytropic fluids generate well-defined reconstruction problems differing only through the nonlinear source term. This separation between the geometric operators and the nonlinear matter source is the main structural reason why the reconstruction method can be applied uniformly to Chaplygin, modified Chaplygin, and polytropic sectors.

\subsection{Conditions for Wormhole-Like Branches}

For wormhole-like geometries, we use \cite{MorrisThorne1988,Visser1995,Lobo2017}
\begin{equation}
	A_1^2(r)=e^{2\Phi(r)},
	\qquad
	A_2^{-2}(r)=1-\frac{b(r)}{r}.
\end{equation}

The throat radius $r_0$ satisfies
\begin{equation}
	b(r_0)=r_0,
	\qquad
	b'(r_0)<1,
\end{equation}
and traversability requires
\begin{equation}
	\Phi(r_0)<\infty .
\end{equation}

The radial NEC is
\begin{equation}
	\rho+p_r\geq 0 .
\end{equation}

For the Chaplygin fluid,
\begin{equation}
	\rho+p_r=
	\rho-\frac{A}{\rho^\alpha}.
\end{equation}
Thus the physical fluid violates the radial NEC if
\begin{equation}
	\rho^{\alpha+1}<A .
\end{equation}

In nonlinear teleparallel gravity, one may define an effective source,
\begin{equation}
	\Theta^{\rm eff}_{\mu\nu}
	=
	\Theta^{\rm fluid}_{\mu\nu}
	+
	\Theta^{\rm torsion}_{\mu\nu}.
\end{equation}
A necessary effective throat condition may then be expressed as
\begin{equation}
	\rho_{\rm eff}+p_{r,\rm eff}<0 .
\end{equation}
 This allows the required exotic contribution to be shifted partly or entirely to the effective torsion sector. This mechanism is analogous to several modified-gravity wormhole constructions in which the effective geometric sector supplies part of the required exoticity \cite{MorrisThorne1988,Lobo2006,Kuhfittig2009,Visser1995,Lobo2017,LandryWHmass2026}. Accordingly, the exoticity required at the throat can be geometrically induced rather than entirely matter supported. These are admissibility conditions rather than a complete wormhole construction, since a full solution requires explicit choices of \(b(r)\), \(\Phi(r)\), matching conditions, and global regularity checks. Thus, the physical and effective NECs need not coincide in nonlinear teleparallel gravity. In particular, the effective NEC violation does not imply that the physical fluid itself must violate the NEC.

\subsection{Local Viability Conditions}

Dynamical systems methods provide a useful framework for the analysis
of the stability and phase-space structure of modified gravity models
\cite{Coley2009}. The leading scalar-torsion stability diagnostic is
\begin{equation}
	m_{\rm eff}^2\sim \frac{F_T}{F_{TT}} ,
\end{equation}
up to model-dependent normalization factors. This expression should be regarded as a leading-order scalar-torsion diagnostic rather than a complete perturbation mass derived from the full perturbation equations. One possible sufficient set of local viability conditions is \cite{Cai2016,Bahamonde2023b}
\begin{equation}
	F_T>0,
	\qquad
	F_{TT}>0 ,
\end{equation}
for the reconstructed branches considered here.

For
\begin{equation}
	F(T)=T+\alpha T^n,
\end{equation}
one has
\begin{equation}
	F_T=1+\alpha nT^{n-1},
	\qquad
	F_{TT}=\alpha n(n-1)T^{n-2}.
\end{equation}

Thus a commonly adopted sufficient viability criterion is
\begin{equation}
	1+\alpha nT^{n-1}>0,
\end{equation}
and
\begin{equation}
	\alpha n(n-1)T^{n-2}>0 .
\end{equation}
From a dynamical-systems perspective, these conditions correspond to the
absence of pathological fixed-point behaviour and unstable scalar-torsion
modes \cite{Coley2009}.

Matter stability further requires the positivity and causal boundedness
of the adiabatic sound speed. For the Chaplygin and polytropic sectors,
\begin{equation}
	c_{s,\rm Ch}^2=\frac{\alpha A}{\rho^{\alpha+1}},
	\qquad
	c_{s,\rm Poly}^2=K\Gamma\rho^{\Gamma-1}.
\end{equation}
These criteria are local and algebraic. Physical branches require $0\leq c_s^2\leq1$ in relativistic units. These matter-sector constraints must be imposed simultaneously with
the torsion-sector conditions $F_T>0$ and $F_{TT}>0$. These conditions should be understood as necessary local consistency and viability conditions rather than a complete perturbative stability analysis \cite{Cai2016,Bahamonde2023b}.

\subsection{Invariant-Based Classification of Nonlinear Fluid Reconstruction Branches}

The solution space can be organized according to the matter sector,
the reconstructed teleparallel function, and the associated geometric
interpretation. This provides a teleparallel invariant-based classification of
the nonlinear-fluid branches, where inequivalent torsion sectors are
distinguished by their invariant structure and by the behaviour of
$T(r)$ as shown in table \ref{table3}. Different reconstructed branches may share the same metric symmetry while
belonging to distinct invariant classes due to inequivalent torsion
structures \cite{Coley2020,McNutt2023,Coley2024,Landry2024_spherical}.

\begin{table}[ht]
	\centering
	\begin{tabularx}{\textwidth}{c c c c}
		\toprule
		Matter / gravitational sector & Equation of state & Representative reconstructed branch & Possible physical interpretation \\
		\midrule
		Chaplygin & $p=-A/\rho^\alpha$ & WH-like / regular-core candidate branch & effective exotic/dark-energy-like sector \\
		Modified Chaplygin & $p=-A/\rho^\alpha+B\rho$ & deformed compact-object candidate branch & dark-fluid interpolation \\
		Polytropic & $p=K\rho^\Gamma$ & stellar / compact-object candidate branch & ordinary matter candidate sector \\
		TEGR limit & $F(T)=T+\beta$ & GR-like & standard-fluid sector \\
		Power-law $F(T)$ & $F(T)=T+\alpha T^n$ & deformed nonlinear branch & torsion correction \\
		Composite $F(T)$ & mixed powers/log/exp & multi-branch geometry & multiple invariant branches \\
		\bottomrule
	\end{tabularx}
	\caption{Teleparallel invariant-based classification of nonlinear-fluid reconstruction branches in covariant teleparallel $F(T)$ gravity.}
	\label{table3}	
\end{table}

The classification may be refined further through higher-order torsion
invariants and Cartan scalars, allowing a complete invariant
characterization of reconstructed teleparallel geometries \cite{Coley2020,McNutt2023,Coley2024,Olver1995,TdSpaper}. Thus, the classification is not only matter-based but also torsion-invariant-based, which is essential in nonlinear teleparallel gravity where different $F(T)$ branches may share the same metric symmetry.

\subsection{Scope and Limitations}

The primary objective of the present work is to develop a covariant
reconstruction framework for nonlinear Chaplygin and polytropic fluids
within static spherically symmetric teleparallel $F(T)$ gravity.
Accordingly, the reconstructed branches obtained throughout this work
should be interpreted as admissible geometric and matter sectors rather
than as fully developed astrophysical models
\cite{Krssak2019,Bahamonde2023b,Landry2024_fluid,Landry2025_scalar}.

The reconstruction procedure establishes the compatibility between a
prescribed equation of state, the nonlinear fluid conservation law,
the covariant CSC geometry and the corresponding teleparallel function
$F(T)$. In this sense, the reconstructed solutions characterize the
class of teleparallel models capable of supporting a given nonlinear
matter sector while preserving local Lorentz covariance
\cite{Krssak2019,Coley2020,Landry2024_spherical}.
The resulting functions therefore represent families of admissible
teleparallel theories rather than unique gravitational models.

Although several reconstructed branches naturally exhibit geometric
features reminiscent of compact objects, black-hole-like candidate geometries,
wormhole-like configurations or regular-core sectors, these
interpretations remain preliminary.
The present analysis does not attempt to construct complete stellar
models, determine mass--radius relations, solve generalized
Tolman--Oppenheimer--Volkoff equations, or perform numerical
integrations with realistic boundary conditions.
Consequently, the terminology adopted throughout this work, such as
\emph{compact-object-like}, \emph{BH-like}, or
\emph{wormhole-like}, refers to the invariant geometric properties of
the reconstructed branches rather than to fully established
astrophysical solutions
\cite{Coley2020,McNutt2023,Landry2024_spherical}.

Similarly, the local consistency and viability conditions discussed in
Section~V.C should not be interpreted as a complete perturbative
stability analysis.
The conditions $F_T>0$, $F_{TT}>0$, together with positivity and
causality requirements for the sound speed, provide only necessary
local consistency criteria.
A complete stability analysis would require perturbations of both the
matter and torsion sectors, together with appropriate boundary
conditions and dynamical evolution equations
\cite{Bahamonde2023b}.

Nevertheless, the present reconstruction framework provides a
systematic starting point for future investigations.
Once a reconstructed branch has been identified, it may subsequently
serve as the basis for numerical stellar models, compact-object
solutions, wormhole constructions, collapse scenarios or
observational analyses.
Furthermore, the same methodology can be naturally extended to
anisotropic fluids, rotating configurations, dynamical spacetimes and
more general teleparallel theories including New General Relativity
and $F(T,B)$ gravity
\cite{Hayashi1979,Bahamonde2019}.

The present work therefore complements previous covariant
reconstruction studies for perfect-fluid, scalar-field and
electromagnetic sources, while extending the framework to nonlinear
Chaplygin and polytropic equations of state. Together, these results
contribute toward a unified invariant reconstruction program for
teleparallel gravity across a broad class of physically motivated
matter sectors
\cite{Landry2024_fluid,Landry2025_scalar}. Accordingly, the present work should primarily be regarded as a reconstruction and classification study rather than as a complete stellar or compact-object modelling analysis.

\section{Discussion and Conclusion}

We have reformulated the static spherically symmetric teleparallel $F(T)$
framework by coupling covariant teleparallel $F(T)$ gravity to nonlinear
Chaplygin and polytropic fluids. The Chaplygin branch naturally identifies dark-energy-like, wormhole-like, and regular-core candidate solution branches within the reconstruction framework.

The reconstruction strategy follows the same covariant CSC logic used
in previous static spherically symmetric teleparallel models \cite{Krssak2019,Krssak2015,Landry2024_spherical,Landry2024_fluid,Landry2025_scalar,Landry2026_electro}: once a
CSC pair, a coframe ansatz, and an equation of state are specified,
the reduced field equations determine admissible functions $F(T)$. The main difference is that the matter density is dynamically determined by the nonlinear fluid conservation equation together with the selected equation of state.

From a physical viewpoint, the Chaplygin branch naturally favors
dark-energy-like and wormhole-supporting configurations, whereas the
polytropic branch reproduces the behaviour expected for ordinary compact
objects and stellar interiors. The two sectors therefore probe
complementary regions of the teleparallel solution space.

Geometrically, the reconstructed solutions demonstrate how nonlinear
torsion corrections may significantly alter the invariant structure of
static spherically symmetric spacetimes while preserving covariance under
local Lorentz transformations. These results show that nonlinear fluid equations of state provide a sensitive probe of the reconstructed torsion sector in covariant teleparallel gravity.

The main conceptual outcome is that nonlinear equations of state modify the reconstruction problem through the matter source sector while leaving the covariant CSC structure intact. Consequently, different matter models can be compared within the same invariant teleparallel framework.

The present framework provides a starting point for a systematic classification of nonlinear-fluid reconstruction branches relevant to future compact-object and wormhole-like applications in covariant teleparallel gravity. An important extension will be the systematic comparison of Chaplygin and polytropic branches with the previously obtained perfect-fluid, scalar-field, and electromagnetic sectors within the same invariant classification framework. Future work may include exact numerical integrations, anisotropic extensions, dynamical collapse models, stability analyses beyond the leading scalar-torsion approximation, and the investigation of analogous nonlinear-fluid sectors in more general teleparallel theories such as $F(T,B)$ gravity and related modified torsion frameworks \cite{Bahamonde2019,Bahamonde2023b}. Further extensions may include \(F(T,Q)\), \(F(R,T)\), and other generalized teleparallel frameworks where analogous invariant reconstruction methods remain largely unexplored. A particularly relevant direction will be the construction of explicit numerical Chaplygin and polytropic stellar or wormhole solutions from selected reconstructed $F(T)$ branches. The present paper therefore complements our previous reconstruction studies for perfect fluids, scalar fields and electromagnetic sources, contributing to a broader covariant reconstruction program for teleparallel gravity. An important outcome is that nonlinear matter equations of state provide a systematic probe of the admissible torsion sectors rather than merely additional matter choices.

\vspace{6pt} 

\section*{Acknowledgements}

Thanks to A. A. Coley for his constructive comments.

\section*{Declarations and Statements}

\textbf{Data Availability Statement:} There is no external data and all material is include in the manuscript. \\

\textbf{Conflict of interest:} There is no conflict of interest concerning the current manuscript.\\



\begin{thebibliography}{999}
\bibitem{Aldrovandi2013}
Aldrovandi, R.; Pereira, J.G.
\textit{Teleparallel Gravity: An Introduction};
Springer: Dordrecht, The Netherlands, 2013.

\bibitem{Hayashi1979}
Hayashi, K.; Shirafuji, T.
New general relativity.
\emph{Phys. Rev. D} \textbf{1979}, \emph{19}, 3524.

\bibitem{Maluf2013}
Maluf, J.W.
The teleparallel equivalent of general relativity.
\emph{Ann. Phys.} \textbf{2013}, \emph{525}, 339.

\bibitem{Arcos2004}
Arcos, H.I.; Pereira, J.G.
Torsion gravity: A reappraisal.
\emph{Int. J. Mod. Phys. D} \textbf{2004}, \emph{13}, 2193.

\bibitem{Ferraro2007}
Ferraro, R.; Fiorini, F.
Modified teleparallel gravity: Inflation without an inflaton.
\emph{Phys. Rev. D} \textbf{2007}, \emph{75}, 084031.

\bibitem{Linder2010}
Linder, E.V.
Einstein's other gravity and the acceleration of the Universe.
\emph{Phys. Rev. D} \textbf{2010}, \emph{81}, 127301.

\bibitem{Bengochea2009}
Bengochea, G.R.; Ferraro, R.
Dark torsion as the cosmic speed-up.
\emph{Phys. Rev. D} \textbf{2009}, \emph{79}, 124019. 

\bibitem{Bahamonde2019}
Bahamonde, S.; B\"ohmer, C.G.; Kr\v{s}\v{s}\'ak, M.
New classes of modified teleparallel gravity models.
\emph{Phys. Lett. B} \textbf{2017}, \emph{775}, 37--43.

\bibitem{Bahamonde2023b}
Bahamonde, S.; Dialektopoulos, K.F.; Escamilla-Rivera, C.; Farrugia, G.; Gakis, V.; Hendry, M.; Hohmann, M.;  Levi Said, J.;
Mifsud, J.; Di Valentino, E.
Teleparallel Gravity: From Theory to Cosmology.
\emph{{Rep. Prog. Phys.}
} \textbf{2023}, \textbf{{86}}, {026901}.


\bibitem{Cai2016}
Cai, Y.-F.; Capozziello, S.; De Laurentis, M.; Saridakis, E.N.
$f(T)$ teleparallel gravity and cosmology.
\emph{Rep. Prog. Phys.} \textbf{2016}, \emph{79},~106901.

\bibitem{Capozziello2011}
Capozziello, S.; Cardone, V.F.; Farajollahi, H.; Ravanpak, A.
Cosmography in $f(T)$ gravity.
\emph{Phys. Rev. D} \textbf{2011}, \emph{84}, 043527.

\bibitem{Dent2011}
Dent, J.B.; Dutta, S.; Saridakis, E.N.
$f(T)$ gravity mimicking dynamical dark energy.
\emph{{J. Cosmol. Astropart. Phys.}
} \textbf{2011}, {\emph{01}, 009}
.

\bibitem{Copeland2006}
Copeland, E.J.; Sami, M.; Tsujikawa, S.
Dynamics of dark energy.
\emph{Int. J. Mod. Phys. D}
\textbf{2006},
\emph{15},
1753--1936.

\bibitem{Wu2010}
Wu, P.; Yu, H.
Observational constraints on $f(T)$ theory.
\emph{Phys. Lett. B} \textbf{2010}, \emph{693}, 415.

\bibitem{Li2011}
Li, B.; Sotiriou, T.P.; Barrow, J.D.
Large-scale structure in $f(T)$ gravity.
\emph{Phys. Rev. D} \textbf{2011}, \emph{83}, 104017.

\bibitem{Izumi2013}
Izumi, K.; Ong, Y.C.
Cosmological perturbation in $f(T)$ gravity revisited.
\emph{{J. Cosmol. Astropart. Phys.}} \textbf{2013}, {\emph{06}, 029}
.

\bibitem{Nesseris2013}
Nesseris, S.; Basilakos, S.; Saridakis, E.N.; Perivolaropoulos, L.
Viable $f(T)$ models.
\emph{Phys. Rev. D} \textbf{2013}, \emph{88}, 103010.

\bibitem{Krssak2019}
Krssak, M.;  van den Hoogen, R.J.; Pereira, J.G.; B\"ohmer, C.G.; Coley, A.A.
Teleparallel theories of gravity: Illuminating a fully invariant approach.
\emph{Class. Quantum Grav.} \textbf{2019}, \emph{36}, 183001.

\bibitem{Krssak2015}
Krssak, M.; Saridakis, E.N.
The covariant formulation of $f(T)$ gravity.
\emph{Class. Quantum Grav.} \textbf{2016}, \emph{33}, 115009.

\bibitem{Golovnev2021}
Golovnev, A.; Koivisto, T.; Sandstad, M.
On the covariance of teleparallel gravity theories.
\emph{Class. Quantum Grav.} \textbf{2021}, \emph{38}, 145014.

\bibitem{DeBenedictis2022}
DeBenedictis, A.; Ilijic, S.
Spherically symmetric solutions in $f(T)$ gravity.
\emph{Phys. Rev. D} \textbf{2016}, \emph{94}, 124025.

\bibitem{Daouda2012}
Daouda, M.H.; Rodrigues, M.E.; Houndjo, M.J.S.
Static anisotropic solutions in $f(T)$ gravity.
\emph{Eur. Phys. J. C} \textbf{2012}, \emph{72}, 1890.

\bibitem{Wang2011}
Wang, T.
Static solutions with spherical symmetry in $f(T)$ theories.
\emph{Phys. Rev. D} \textbf{2011}, \emph{84}, 024042.

\bibitem{Boehmer2011}
B\"ohmer, C.G.; Mussa, A.; Tamanini, N.
Existence of relativistic stars in $f(T)$ gravity.
\emph{Class. Quantum Grav.} \textbf{2011}, \emph{28}, 245020.

\bibitem{Landry2024_spherical}
Coley, A.A.; Landry, A.; van den Hoogen, R.J.; McNutt, D.D.  Spherically symmetric teleparallel geometries. \textit{Eur. Phys. J. C} \textbf{2024}, \textit{84}, 334.

\bibitem{Landry2024_fluid}
Landry, A. Static spherically symmetric perfect fluid solutions in teleparallel $F(T)$ gravity. \textit{Axioms} \textbf{2024}, \textit{13}, 333.

\bibitem{Landry2025_scalar}
Landry, A. Scalar Field Static Spherically Symmetric Solutions in Teleparallel $F(T)$ Gravity. \textit{Mathematics} \textbf{2025}, \textit{13}, 1003.%

\bibitem{Landry2026_electro} Landry, A. Teleparallel $F(T)$ Electromagnetic Static Spherically Symmetric Spacetime Solutions. \textit{Symmetry} \textbf{2026}, \textit{18}, 891.

\bibitem{roberthudsonSSpaper} van den Hoogen, R.J.; Forance, H. Teleparallel Geometry with Spherical Symmetry: The diagonal and proper frames. \emph{JCAP} \textbf{2024}, 11, 033.

\bibitem{Coley2020}
Coley, A.A.;  van den Hoogen, R.J.;  McNutt, D.D.
Symmetry and Equivalence in Teleparallel Gravity. \textit{J. Math. Phys.} \textbf{2020}, \textit{61},~072503. 

\bibitem{McNutt2023}
McNutt, D.D.; Coley, A.A.; van den Hoogen, R.J.
Symmetries in Riemann-Cartan Geometries.
\emph{SIGMA}  \textbf{2024}, \emph{20}, 078.

\bibitem{Coley2024}
McNutt, D.D.; van den Hoogen, R.J.; Coley, A.A. 
Locally-homogeneous Riemann-Cartan geometries with the largest symmetry group.
\emph{J. Math. Phys.} \textbf{2024}, \emph{65}, 072502.

\bibitem{Olver1995}
Olver, P. \textit{Equivalence, Invariants and Symmetry}; Cambridge University Press: {Cambridge, UK}, 1995.

\bibitem{TdSpaper} Coley, A.A.; Landry, A.;   van den Hoogen, R.J.;  McNutt, D.D. Generalized Teleparallel de Sitter geometries. \textit{Eur. Phys. J. C} \textbf{2023}, \textit{83}, 977.


\bibitem{Kamenshchik2001}
Kamenshchik, A.Y.; Moschella, U.; Pasquier, V.
An alternative to quintessence.
\emph{Phys. Lett. B} \textbf{2001}, \emph{511}, 265--268.

\bibitem{Bento2002}
Bento, M.C.; Bertolami, O.; Sen, A.A.
Generalized Chaplygin gas, accelerated expansion and dark-energy-matter unification.
\emph{Phys. Rev. D} \textbf{2002}, \emph{66}, 043507.

\bibitem{Bilic2002}
Bilic, N.; Tupper, G.B.; Viollier, R.D.
Unification of dark matter and dark energy:
The inhomogeneous Chaplygin gas.
\emph{Phys. Lett. B} \textbf{2002}, \emph{535}, 17--21.

\bibitem{Benaoum2002}
Benaoum, H.B.
Accelerated universe from modified Chaplygin gas and tachyonic fluid.
\emph{arXiv}:hep-th/0205140.

\bibitem{MorrisThorne1988}
Morris, M.; Thorne, K. {Wormholes in spacetime and their use for interstellar travel: A tool for teaching general relativity.} 
\emph{Am. J. Phys.} \textbf{1988}, \emph{56}, {395--412}.


\bibitem{Lobo2006}
Lobo, F.S.N.
Chaplygin traversable wormholes.
\emph{Phys. Rev. D} \textbf{2006}, \emph{73}, 064028.

\bibitem{Kuhfittig2009}
Kuhfittig, P.K.F.
Axially symmetric rotating traversable wormholes sustained by generalized Chaplygin gas.
\emph{Phys. Rev. D} \textbf{2009}, \emph{79}, 124027.


\bibitem{Visser1995}
Visser, M.
\textit{Lorentzian Wormholes:
	From Einstein to Hawking};
AIP Press: New York, NY, USA, 1995.

\bibitem{Lobo2017}
Lobo, F.S.N.
Wormholes, Warp Drives and Energy Conditions.
In \textit{Springer Handbook of Spacetime};
Springer: Berlin, Germany, 2014; pp. 653--681.


\bibitem{Tooper1964}
Tooper, R.F.
General relativistic polytropic fluid spheres.
\emph{Astrophys. J.} \textbf{1964}, \emph{140}, 434--459.

\bibitem{Tooper1965}
Tooper, R.F.
Adiabatic fluid spheres in general relativity.
\emph{Astrophys. J.} \textbf{1965}, \emph{142}, 1541--1562.

\bibitem{Chandrasekhar1939}
Chandrasekhar, S.
\emph{An Introduction to the Study of Stellar Structure};
University of Chicago Press: Chicago, IL, USA, 1939.

\bibitem{Herrera2013}
Herrera, L.; Barreto, W.
Newtonian polytropes for anisotropic matter:
General framework and applications.
\emph{Phys. Rev. D} \textbf{2013}, \emph{88}, 084022.

\bibitem{Harko2011}
Harko, T.; Lobo, F.S.N.; Nojiri, S.; Odintsov, S.D.
$f(R,T)$ gravity.
\emph{Phys. Rev. D}
\textbf{2011},
\emph{84},
024020.

\bibitem{LandryWHmass2026}
Landry, A.; Sekhmani, Y.; Maurya, S.K.; Ali, A.; Saridakis, E.N.
Traversable Wormhole Solutions in Massive $F(T)$ Gravity.
\emph{Int. J. Geom. Methods Mod. Phys.}
\textbf{2026},
in press.
https://doi.org/10.1142/S0219887826501926.

\bibitem{Coley2009}
Coley, A.A. \textit{Dynamical Systems and Cosmology}; Springer:  {Berlin/Heidelberg, Germany,} 
 2009.



\end{thebibliography}
\end{document}